\documentclass[11pt]{article}
\usepackage{latexsym}  
\usepackage{amssymb}
\usepackage{graphicx}
\usepackage{amsmath}

\begin{document}


\begin{center}

{\Large\bf Parameter-Independent Quark Mass Relation } \\[2mm]

{\Large\bf in the U(3)$\times$U(3)$'$ Model }

\vspace{4mm}

{\bf Yoshio Koide$^a$ and Hiroyuki Nishiura$^b$}

${}^a$ {\it Department of Physics, Osaka University, 
Toyonaka, Osaka 560-0043, Japan} \\
{\it E-mail address: koide@kuno-g.phys.sci.osaka-u.ac.jp}

${}^b$ {\it Faculty of Information Science and Technology, 
Osaka Institute of Technology, 
Hirakata, Osaka 573-0196, Japan}\\
{\it E-mail address: hiroyuki.nishiura@oit.ac.jp}

\date{\today}
\end{center}

\vspace{3mm}

\begin{abstract}
Recently, we have proposed a quark mass matrix model
based on U(3)$\times$U(3)$'$ family symmetry,
in which up- and down-quark mass matrices $M_u$ and $M_d$   
are described only by complex parameters 
$a_u $ and $a_d $, respectively. 
When we use charged lepton masses as additional input values,
we can successfully obtain predictions for quark masses 
and Cabibbo-Kobayashi-Maskawa mixing. 
Since we have only one complex parameter $a_q$ for 
each mass matrix  $M_q$, 
we can obtain a parameter-independent mass relation  by using three equations for ${\rm Tr}[H_q]$, 
${\rm Tr}[H_q H_q]$ and ${\rm det}H_q$, where 
$H_q \equiv M_q M_q^\dagger$ ($q=u, d$). 
In this paper, we investigate its parameter-independent feature of the quark mass relation in the model.

\end{abstract}

PCAC numbers:  
  11.30.Hv, 
  12.15.Ff, 
  12.60.-i, 

\vspace{3mm}

{\large\bf 1  \  Introduction} 
 
Recently, we have proposed a quark mass matrix model\cite{KN_PRD15}
based on U(3)$\times$U(3)$'$ symmetry, 
in which mass matrices for up-quarks, down-quarks, charged leptons, and neutrinos,  
$M_f$ ($f=u,d,e, \nu$) ,  
are described  respectively with only one complex parameters $a_f$ by
$$
(M_f)_i^{\ j} = m_{0f} (\Phi_f )_i^{\ \alpha} 
(S_f^{-1})_\alpha^{\ \beta} (\bar{\Phi}_f)_\beta^{\ j}.
\eqno(1.1)
$$
Here $\Phi_f$ and $S_f$ are vacuum expectation values (VEVs) matrices.  
The  indexes $i, j=1,2,3$ are ones of U(3) family and 
$\alpha, \beta =1,2,3$ are indexes of U(3)$'$ family. 
Although  $\Phi_f$ and $S_f$ 
have a dimension of ''mass", we put the factor $m_{0f}$ with a 
dimension of mass in Eq.(1.1), since we treat those as dimensionless 
quantities as seen in (1.2),  (1.3) and (1.5) later.

In (1.1),  $M_\nu$ is a Dirac neutrino mass matrix. 
Although we consider that the observed neutrinos are 
Majorana neutrinos and the Majorana neutrino mass 
matrix is given by a similar mechanism 
\cite{KN_PRD15,KN_JMPA17} to the so-called 
neutrino seesaw mechanism \cite{nu_seesaw}, 
we do not discuss the structure of $M_\nu$ in the 
present paper because the purpose of the present 
paper is to discuss the quark mass relation.
  
We define structure of the matrix $\Phi_f$ as an dimensionless 
expression 
$$
\Phi_f = \Phi_0 P_f , 
\eqno(1.2)
$$
where
$$
\Phi_0 ={\rm diag} (z_1, z_2, z_3 ) , 
\eqno(1.3)
$$
$$
P_f ={\rm diag} (e^{i\phi_{f1}}, e^{i\phi_{f2}},  e^{i\phi_{f3}}) . 
\eqno(1.4)
$$

Since we consider that the U(3)$'$ is broken into 
a discrete symmetry S$_3$,
the matrix  $ (S_f^{-1})$ is given by
$$
(S_f)^{-1} =({\bf 1} + a_f X) = ({\bf 1} + b_f X )^{-1} ,
\eqno(1.5)
$$
where 
$$
{\bf 1} = \left( \begin{array}{ccc}
 1 & 0  & 0 \\
 0 & 1 & 0 \\
 0 & 0 & 1 
\end{array} \right), \ \ \ 
X = \frac{1}{3}  \left( 
\begin{array}{ccc} 
1 & 1  & 1 \\
 1 & 1 & 1 \\
 1 & 1 & 1 \end{array} \right) , 
\eqno(1.6)
$$
and $a_f$ is a complex parameter: 
$$
a_f = - \frac{b_f}{1+ b_f} .
\eqno(1.7)
$$
Only for the charged lepton mass matrix $M_e$, 
the parameter $a_e$ is given by $a_e=0$, so that 
the mass matrix $M_e$ is given by 
$$
M_e = m_{0e} \Phi_e \Phi_e^\dagger = m_{0e} \Phi_0 \Phi_0 ,
\eqno(1.8)
$$
where we can put $P_e = {\bf 1}$ without losing a generality. 
Namely, we take $S_e = {\bf 1}$ only for $f=e$. 
Therefore, the parameters $z_i$ ($i=1,2,3$) are given by
$$
z_i = \sqrt{ \frac{m_{e_i}}{ m_{e1} +m_{e2} + m_{e3}} } ,
\eqno(1.9)
$$
where $( m_{e1}, m_{e2}, m_{e3}) = (m_e, m_\mu, m_\tau)$. 
Here, as the input values $ (m_e, m_\mu, m_\tau)$, the running  
mass values of the charged leptons at a scale $\mu = m_Z$,
$(m_e, m_\mu, m_\tau) = (0.000486849\ {\rm GeV}, 
0.102751\ {\rm GeV}, 1.7467\ {\rm GeV})$, are used, not the 
pole masses,  because the predicted quark mass values are 
calculated at the scale $\mu = m_Z$.  
(The study of the quark mass matrix (1.1) with the form 
(1.5) has been substantially done in Ref.\cite{YK-HF_ZPC96}
although the model has been based on U(3)-family symmetry,
not U(3)$\times$U(3)$'$.)

In this model, when we choose suitable values of the 
complex parameters $a_q $ ($q = u, d$) together with 
additional input values, $(m_e, m_\mu, m_\tau)$,  
we can successfully obtain \cite{KN_PRD15} predictions 
for quark masses and Cabibbo-Kobayashi-Maskawa (CKM)mixing
\cite{CKM}. 
However, so far, it is not clear whether the successful 
parameter fitting is unique or not,  
and that there are another good parameter solutions or not. 

In order to settle  these questions, 
parameter-independent mass relations are usefull, which can be obtained in this model ;
We have three independent equations for the each mass matrix 
$M_q$ ($q=u, d$), while we have only one complex 
parameter $a_q$, therefore we can obtain one mass relation.    
In this paper, we investigate such the parameter-independent 
mass relation in the U(3)$\times$U(3)$'$ model.  
At present, the observed quark mass values, especially, for 
the first generation quarks have considerably large error,
i.e. $m_u = 1.27^{+0.50}_{-0.42}$ MeV and $m_d= 2.90^{+1.24}_{-1.19}$ 
MeV at $\mu = m_Z$ \cite{q-mass}.  
By obtaining such a parameter-independent quark mass relation, 
we check whether the U(3)$\times$U(3)$'$ model is 
reasonable or not and what values of $m_u$ and $m_d$ are
acceptable to  the U(3)$\times$U(3)$'$ model. 
\\


{\large\bf 2  \  Brief review of the U(3)$\times$U(3)$'$ model  }

In our model based on U(3)$\times$U(3)$'$ symmetry\cite{KN_PRD15, KN_JMPA17}, 
we consider hypothetical fermions $F_\alpha$ 
($\alpha=1,2,3$), which belong to $({\bf 1}, {\bf 3})$ of
 U(3)$\times$U(3)$'$, 
in addition to quarks and leptons $f_i$ ($i=1,2,3$) 
which belong to  $({\bf 3}, {\bf 1})$.   

We assume that the VEV form (1.1) 
originates from the following $6\times 6$ mass matrix model:
$$
(\bar{f}_L^i \ \ \bar{F}_L^\alpha ) 
\left(
\begin{array}{cc}
(0)_i^{\ j}  &  (\Phi_f)_i^{\ \beta}  \\
(\bar{\Phi}_{f})_\alpha^{\ j} & -(S_f)_\alpha^{\ \beta} 
\end{array} \right) 
 \left(
\begin{array}{c}
f_{Rj} \\
F_{R\beta}
\end{array} 
\right) .
\eqno(2.1)
$$
Here, $F_{L(R)}$ are heavy fermions with 
$({\bf 1}, {\bf 1}, {\bf 3})$ of
SU(2)$_L \times$U(3)$\times$U(3)$'$.
On the other hand, 
$f_R$ are right-handed quarks and leptons,
$f_R= (u, d, \nu, e^-)_R$, while
$f_L$ are not physical fields.  
They are given by the following combinations: 
$$
f_L \equiv (f_u, f_d, f_\nu, f_e)_L \equiv 
\left( \frac{1}{\Lambda_H} H_u^\dagger q_L, 
\frac{1}{\Lambda_H} H_d^\dagger q_L,
\frac{1}{\Lambda_H} H_u^\dagger \ell_L,
\frac{1}{\Lambda_H} H_d^\dagger \ell_L \right) ,
\eqno(2.2)
$$
where $\Lambda_H$ is a flavon VEV scale,  and
$$
q_L = \left(
\begin{array}{c}
u_L \\
d_L 
\end{array} \right) , \ \ \ \ 
\ell_L = \left(
\begin{array}{c}
\nu_L \\
e^-_L 
\end{array} \right) , \ \ \ \ 
H_u = \left(
\begin{array}{c}
H_u^0 \\
H_u^- 
\end{array} \right) , \ \ \ \ 
H_d = \left(
\begin{array}{c}
H_d^+ \\
H_d^0 
\end{array} \right) . 
\eqno(2.3)
$$
In other words, the matrix given in Eq.(2.1) denotes would-be 
Yukawa coupling constants. 

After the U(3) and U(3)$'$ have been completely broken, 
the quarks and leptons are described by the effective 
Hamiltonian
$$
{\cal H}_Y = (\bar{\nu}_L)^i (M_\nu)_i^{\ j}  (\nu_R)_j 
+ (\bar{e}_L)^i (M_e)_i^{\ j} (e_R)_j  
+ y_R (\bar{\nu}_R)^i (Y_R)_{ij} (\nu_R^c)^j 
$$
$$
+  (\bar{u}_L)^i (M_u)_i^{\ j} (u_R)_j 
+ (\bar{d}_L)^i (M_d)_i^{\ j}  (d_R)_j  .
\eqno(2.4)
$$
Note that the quarks and leptons $f_i$ are not U(3) family 
triplet any more in the exact meaning, but they are mixing states 
between the fermions $f$ and $F$.
However, for convenience,  we will still use the index of 
U(3) family for these fermion states. 

By performing a seesaw-like approximation with $\Lambda_2 =O(\langle \Phi_f \rangle)
 \ll \Lambda_1=O(\langle S_f \rangle)$, 
the mass matrix (2.1) leads to the following Dirac mass matrix 
of quarks and leptons:
$$
(M_f)_i^{\ j} \simeq \frac{\langle H_{u/d} 
\rangle}{\Lambda_H} 
 \langle {\Phi}_{f} \rangle_i^{\ \alpha} 
\langle (S_f)^{-1} \rangle_{\alpha}^{\ \beta} 
\langle \bar{\Phi}_{f} \rangle_{\beta}^{\ j} .
\eqno(2.5)
$$
However, 
in this paper, since we interest only in the relative ratios 
of the quark masses in the same sector ($q=u$ or $q=d$), 
the factor  $\langle H_{u/d} \rangle / \Lambda_H $ 
takes a common value, so that  the factor 
$\langle H_{u/d} \rangle / \Lambda_H $ does not play any 
essential role in our study.

As seen in this section, $\Phi_f$ and $S_f$ have a dimension 
of mass.  However, for convenience, hereafter, we use 
a dimensionless  expressions (1.2) and (1.5), and 
define the parameter $m_{0f}$ with a mass dimension
by Eq.(1.1). 
\\ 

{\large\bf 3  \  Three equations for mass relations  }
 
When we define an Hermitian mass matrix
$$
H_q  = M_q M_q^\dagger, 
\eqno(3.1)
$$
the explicit form of $H_q$ is given by
$$
H_q = M_q M_q^\dagger = m_{0q}^2 \Phi_q S_q^{-1} \Phi_q^\dagger 
\Phi_q (S_q^\dagger)^{-1} \Phi_q^\dagger 
$$
$$
=  k_q^{\ 2} P_q D_e^{1/2} ({\bf 1} + a_q X)  D_e ({\bf 1} + a_q^* X)
D_e^{1/2} P_q^{\ \dagger} , 
\eqno(3.2)
$$
where 
$$
k_q = \frac{m_{0q}}{m_{0e}} .
\eqno(3.3)
$$

The Hermitian matrix $H_q$ is diagonalized as 
$$
U_q H_q U_q^\dagger = D_q^{\ 2} \equiv 
{\rm diag}(m_{q1}^{\ 2}, m_{q2}^{\ 2}, m_{q3}^{\ 2} ) .
\eqno(3.4)
$$
Hereafter, for convenience, we define 
$$
(m_1, m_2, m_3) = \frac{1}{k_q} (m_{q1}, m_{q2}, m_{q3}) ,
\eqno(3.5)
$$
and 
$$
\tilde{D}_q \equiv {\rm  diag}(m_{1}, m_{2}, m_{3} ) 
=\frac{1}{k_q} D_q .
\eqno(3.6)
$$

In general, we have the following three equations for the matrix $H_q$:
$$
c_1 \equiv m_{1}^{\ 2} +m_{2}^{\ 2} +m_{3}^{\ 2} = 
{\rm Tr}[\tilde{D}_q^{\ 2}] =\frac{1}{k_q^{\ 2}}  {\rm Tr}[H_q] , 
\eqno(3.7) 
$$
$$
c_2 \equiv m_{1}^{\ 2} m_{2}^{\ 2} + m_{2}^{\ 2} m_{3}^{\ 2} 
+m_{3}^{\ 2} m_{1}^{\ 2} = 
\frac{1}{2} \left( 
({\rm Tr}[\tilde{D}_q^{\ 2}])^2 -{\rm Tr}[\tilde{D}_q^{\ 2} \tilde{D}_q^{\ 2}]
\right)
$$
$$
= \frac{1}{k_q^{\ 4} } \frac{1}{2} 
\left\{
({\rm Tr}[H_q])^2-  {\rm Tr}[H_q H_q]  
\right\} , 
\eqno(3.8)
$$
 and 
$$
c_3 \equiv m_{1}^{\ 2} m_{2}^{\ 2} m_{3}^{\ 2} = 
 {\rm det} \tilde{D}_q^{\ 2} =\frac{1}{k_q^{\ 6} }  {\rm det}H_q . 
\eqno(3.9)
$$
For convenience, hereafter, we denote Tr$[A]$ as $[A]$ simply. 
By using the explicit form (3.2), we obtain  $c_1$, $c_2$ and $c_3$ as follows:
$$
 c_1 =  \left| 1 +\frac{1}{3} a_q \right| ^2 [D_e^2] + \frac{1}{9} |a_q|^2 
([D_e]^2 -[D_e^2] ) ,
\eqno(3.10)
$$
$$
c_2=\frac{1}{2} \left| 1+ \frac{2}{3} a_q \right|^2 ( [D_e^2]^2 -[D_e^4]) 
+\frac{2}{9} |a_q|^2 [D_e] {\rm det} D_e ,
\eqno(3.11)
$$
$$
c_3 = | 1+ a_q|^2 ({\rm det}D_e)^2 =  | 1+ a_q|^2 {\rm det} D_e^2 . 
\eqno(3.12)
$$

Let us define
$$
 Q_1 \equiv \frac{ [D_q^2]}{ [D_e^2]} , \ \ \ 
 Q_2 \equiv \frac{  [D_q^2]^2 -[D_q^4] }{ [D_e^2]^2 -[D_e^4] } , \ \ \ 
Q_3 \equiv \frac{ \det D_q^2}{\det D_e^2} , 
\eqno(3.13) 
$$
and 
$$
L_1 \equiv \frac{[D_e]^2 -[D_e^{\ 2}]}{[D_e^{\ 2}] } , \ \ \ \ 
L_2 \equiv \frac{[D_e]\det D_e }{ [D_e^{\ 2} ]^2 -[D_e^{\ 4}] } .
\eqno(3.14)
$$
then, we obtain the following relations:
$$
Q_1 = \left| 1 +\frac{1}{3} a_q \right| ^2  + \frac{1}{9}|a_q|^2  L_1 , 
\eqno(3.15)
$$
$$
Q_2 = \left| 1+ \frac{2}{3} a_q \right|^2  + \frac{4}{9}  |a_q|^2 L_2 , 
\eqno(3.16)
$$
$$
Q_3 =  | 1+ a_q |^2 .
\eqno(3.17)
$$
Note that $L_1$ and $L_2$ are given only by the charged lepton 
masses, and $Q_1$, $Q_2$ and $Q_3$ are expressed by quark 
masses $(m_1, m_2, m_3)$ after $(m_e, m_\mu, m_\tau)$ are 
substituted.

Finally, by eliminating the parameter $a_q$ from Eqs.(3.15) -(3.17), 
we obtain the mass relation
$$
-b_0 + b_1 (m_{1}^{\ 2} + m_{2}^{\ 2} + m_{3}^{\ 2} ) -
b_2 (m_{1}^{\ 2} m_{2}^{\ 2} + m_{2}^{\ 2} m_{3}^{\ 3} +
m_{3}^{\ 2} m_{1}^{\ 2}) +b_3\, m_{1}^{\ 2} m_{2}^{\ 2} m_{3}^{\ 2} =0 .
\eqno(3.18)
$$
Here the coefficients $b_0$, $b_1$, $b_2$, and $b_3$ are defined by 
$$
b_0 = \left(1+\frac{1}{2} L_1 -4 L_2\right) ,
\eqno(3.19)
$$
$$
 b_1 = 3 (1-2 L_2) \frac{1}{[D_e^{\ 2}]} , 
\eqno(3.20)
$$
$$
 b_2 = 6 \left(1 +\frac{1}{2} L_1\right) \frac{1}{ [D_e^{\ 2}] -[D_e^{\ 4}] } , 
 \eqno(3.21)
 $$
 $$
 b_3 = (1-L_1 + 2 L_2) \frac{1}{\det D_e^{\ 2} } ,
\eqno(3.20)
$$
which are expressed only by the charged lepton masses. 
\\


{\large\bf 4  \  Behavior of $m_1/m_2$ versus $m_2/m_3$  }

 In order to investigate the behavior of   $m_1/m_2$ versus $m_2/m_3$, 
we define parameters
$$
x \equiv \frac{m_2}{m_3} =  \frac{m_{q2}}{m_{q3}} , 
\ \ \ \ \ 
y  \equiv \frac{m_1}{m_2} =  \frac{m_{q1}}{m_{q2}} .
\eqno(4.1)
$$ 
Note that the parameters $x$ and $y$ are independent of 
the value of $k_q$ defined in (3.3).  
Then, since $(m_1, m_2, m_3 )$ are expressed as
$$
(m_1, m_2, m_3 ) = ( x y,  x, 1) m_3 , 
\eqno(4.2)
$$
the relation (3.18) becomes 
$$
- b_0 + b_1 m_3^{\ 2} (1+ x^2 + x^2 y^2)  
- b_2 m_3^{\ 4} (x^2+x^2 y^2 + x^4 y^2)  +  b_3\, m_3^{\ 6}  x^4 y^2 = 0. 
\eqno(4.3)
$$
Therefore, we get a relation $y=f(x)$:
$$
y= \frac{1}{x} \sqrt{ \frac{ (b_1 m_3^{\ 2} - b_2 m_3^{\ 4} ) x^2 
- ( b_0 - b_1 m_3^{\ 2} )}{ ( b_2 m_3^{\ 4} - b_3 m_3^{\ 6} )x^2 
- (b_1 m_3^{\ 2} - b_2 m_3^{\ 4} )  } }.
\eqno(4.4)
$$
As seen in (4.4), the function  $y=f(x)$ has poles at 
$$
x= 0, \ \ \ \ {\rm and} \ \ \ 
x= \pm \sqrt{ \frac{ b_1 m_3^{\ 2} - b_2 m_3^{\ 4} }{
b_2 m_3^{\ 4} - b_3 m_3^{\ 6} } } ,
\eqno(4.5)
$$
and a zero point at
$$
x = \pm \sqrt{ \frac{  b_0 - b_1 m_3^{\ 2} }{ b_1 m_3^{\ 2} 
- b_2 m_3^{\ 4} } } .
\eqno(4.6)
$$

The explicit values $b_0$, $b_1$, $b_2$ and $b_3$ are given by
$$
\begin{array}{l} 
b_0 = 1.04888, \\
b_1 = 0.97489 \ {\rm (GeV)^{-2} }, \\
b_2 = 87.6457 \ {\rm (GeV)^{-4} }, \\
b_3 = 1.16204\times 10^8 \ {\rm (GeV)^{-6}} . 
\end{array}
\eqno(4.7)
$$
Here we have used running mass values $m_e (\mu)$ = 0.000486847 GeV, $m_\mu (\mu)$ = 0.102751
GeV and $m_\tau (\mu)$ = 1.7467 GeV as the charged lepton mass values at $\mu =m_Z$ \cite{q-mass} .

The behavior of $y=f(x)$ is illustrated in Fig.1.
The behavior depends on the input value of $m_3$. 
Note that since 
$$
\frac{b_0}{ b_1} = 1.0758 \ ({\rm GeV})^2 , 
\eqno(4.8)
$$
the factor $(b_0 -b_1 m_3^{\ 2})$ in (4.4) changes the sign according as 
$m_3 > m_{30}$ or $m_3 < m_{30}$,
where
$$
m_{30} \equiv \sqrt{\frac{b_0}{b_1} } = 1.0373 \ {\rm GeV} .
\eqno(4.9)
$$
Hereafter, we call the behavior in the case  $m_3 > m_{30}$ 
as normal type, and the behavior in the case $m_3 < m_{30}$ 
as non-normal type. 

Now, let us compare our parameter-independent results with  
the observed quark mass values in detail.
The observed quark mass values at  at $\mu =m_Z$ \cite{q-mass} are 
as follows: 
$$
\begin{array}{lll}
m_u = 0.00127^{+0.00050}_{-0.00042}\ {\rm GeV}, & 
m_c = 0.619 \pm 0.084\ {\rm GeV}, & 
m_t =171.7 \pm 3.0 \ {\rm GeV}, \\
m_d = 0.00290 ^{+0.00124}_{-0.00119} \ {\rm GeV}, &
m_s = 0.055^{+0.016}_{-0.015}\ {\rm GeV}, &
m_b =2.89 \pm0.09 \ {\rm GeV}. 
\end{array}
\eqno(4.10)
$$
Hereafter, we use the following values as the mass ratios
$m_1/m_2$ and $m_2/m_3$: 
$$
\frac{m_u}{m_c} = 0.00205^{+ 0.00126}_{-0.00084} , \ \ \ \ \ 
\frac{m_c}{m_t} = 0.00361^{+0.00056}_{-0.00054} , 
\eqno(4.11)
$$

$$
\frac{m_d}{m_s} = 0.0527^{+ 0.0508}_{-0.0286} , \ \ \ \ \ 
\frac{m_s}{m_b} = 0.0190^{+0.00063}_{-0.00056} .
\eqno(4.12)
$$

\begin{figure}[t!]
  \includegraphics[width=140mm,clip]{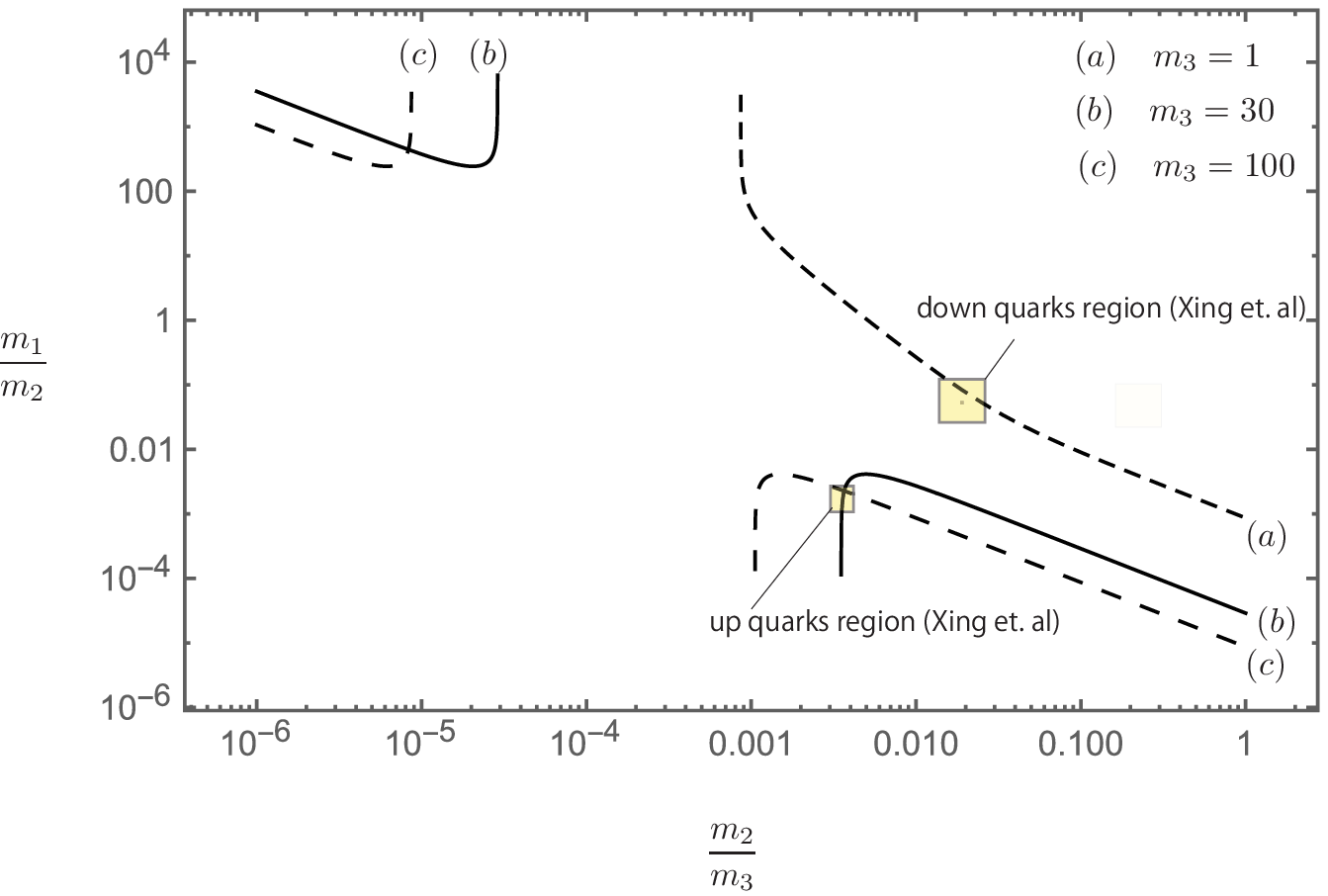}
 \begin{quotation}
Figure 1: \,\,$m_3$ dependence of the behavior of  $m_1/m_2$ 
versus $m_2/m_3$. The curves of the mass relation $y=f(x)$ 
given in Eq.(4.4) are drown in the $(m_2/m_3, m_1/m_2 )$ 
plane for the cases (a) $m_3=1$ GeV,  (b) $m_3=30$ GeV,  
and (c) $m_3=100$ GeV. The shaded square regions are 
correspond to the observed mass ratios  in (4.11) and 
(4.12) for up-quark sector and down-quark sector 
respectively obtained by Xing et. al. 
\end{quotation}
  \label{fig1}
\end{figure}

As seen in Fig.1, the behavior of $m_1/m_2$ in 
the  normal type has a maximum whose value is smaller than
$\sim 10^{-2}$.
On the other hand,
as seen in Eq.(4.12), the observe value of $m_d/m_s$ is
$m_d/m_s \simeq 0.05$.
Therefore, the mass ratios for down-quark sector cannot 
be described by the behavior of the normal type. 
Thus we have the solution for Eqs.(4.11) and (4.12) 
by the behavior of the normal type for the up-quark 
sector, and by the behavior of the non-normal type 
for the down-quark sector. 

\vspace{2mm}

\begin{figure}[h!]
  \includegraphics[width=140mm,clip]{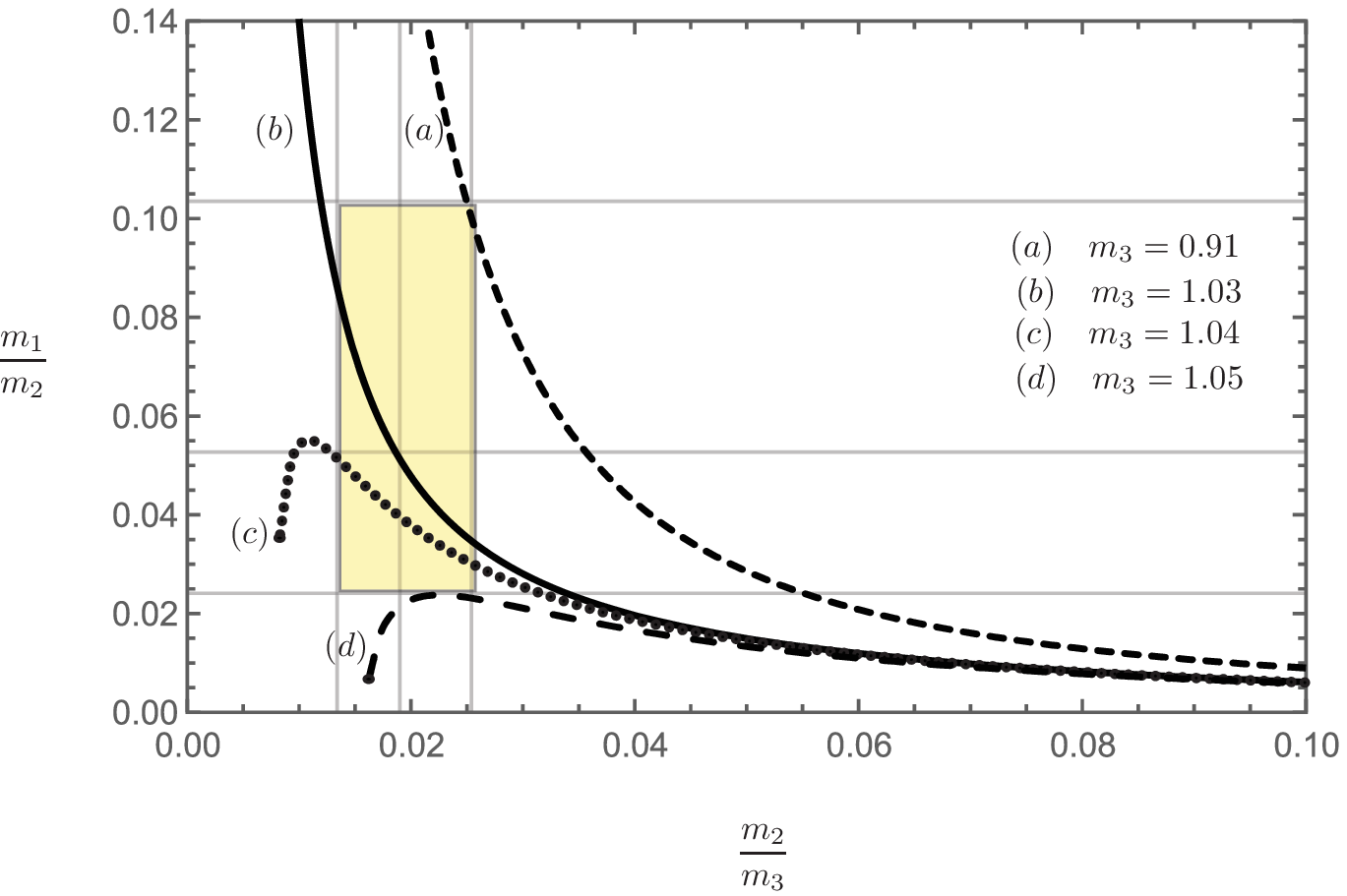}
\begin{quotation}
Figure 2: \,\,The $m_3$ value in the mass relation consistent 
with the observed values $m_d/m_s$ and  $m_s/m_b$ in the 
down-quark  sector. The curves of the mass relation $y=f(x)$ 
in Eq.(4.4) are drown in the $(m_2/m_3, m_1/m_2 )$ plane 
for the cases with (a) $m_3=0.91$ GeV,  (b) $m_3=1.03$ GeV,  
(c) $m_3=1.04$ GeV, and (d) $m_3=1.05$ GeV. The each curve in the case of (c) and (d) has a left end point, in the left  region from which there is no real solutions for $m_1/m_2$. 
The shaded square region is correspond to the observed mass ratios 
in (4.12) for  down-quark sector obtained by Xing et. al. 
\end{quotation}
  \label{fig-down}
\end{figure}

\vspace{2mm}

\noindent{\bf Down-quark sector} 

First, let us see behaviors of the mass ratios 
$(m_1/m_2, m_2/m_3)$ in the down-quark sector. 

As seen in Fig.2, we can determine a value 
$m_3$ from the observed center values in (4.12)  
as $m_3 =1.03$ GeV. 
If we take $m_3 > 1.04$ GeV, the behavior of the 
mass ratios becomes the normal type from 
non-normal type as seen in the curve (c) of Fig.2. 
Furthermore, if we take a larger value $m_3 >1.05$ GeV, then  
the curve is out of the error region as seen in 
the curve (d) of Fig.2.
Similarly, there is no solution of $m_3$ for 
$m_3 < 0.91$ GeV as seen in the curve (a) of Fig.2. 
Thus we obtain 
$$
m_3 = 1.03^{+ 0.02}_{-0.12} \,\, \mbox{GeV}, 
\eqno(4.13)
$$
from the consistency between and the mass relation (4.4) 
and  the observed mass ratios  (4.12).

If we take $m_3=1.03$ GeV, 
we obtain
$$
k_d = \frac{m_b}{m_3} = 2.89, 
\eqno(4.14)
$$
 from the definition (3.5).  
In this choice of the value of  $k_d$, we have 
$$
m_b = k_d m_3 = 2.98^{+ 0.06}_{-0.35} \,\, \mbox{GeV}, 
\eqno(4.15)
$$
which has smaller error bar for upper limit than 
that of the observed $m_b$ in (4.10).
\vspace{2mm}

\begin{figure}[h!]
  \includegraphics[width=140mm,clip]{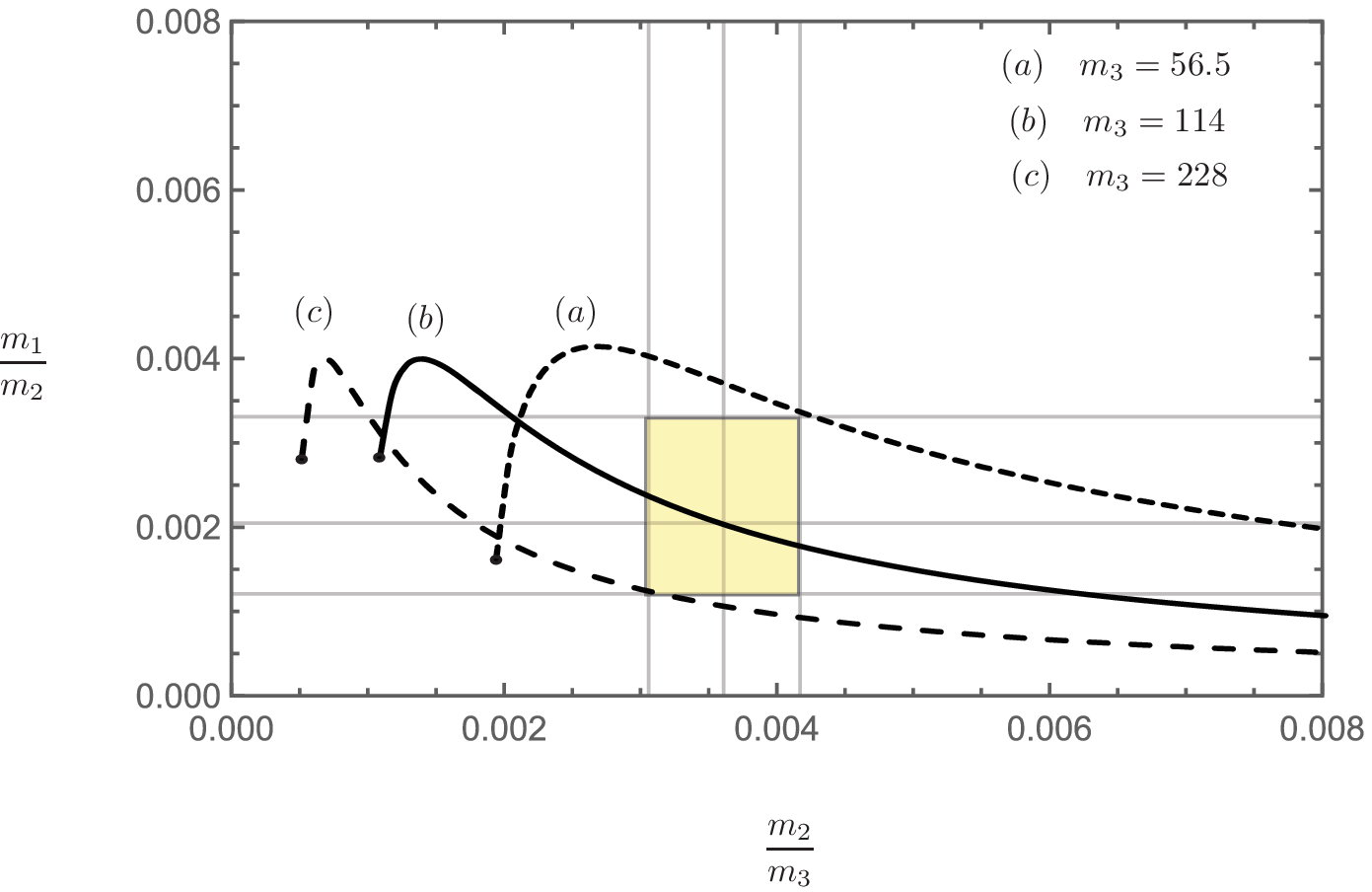}
\begin{quotation}
Figure 3: \,\,The $m_3$ value in the mass relation consistent 
with the observed values  $m_u/m_c$ and $m_c/m_t$ in 
the up-quark  sector. The curves of the mass relation 
$y=f(x)$ in Eq.(4.4) are drown in the $(m_2/m_3, m_1/m_2 )$ 
plane for the cases with (a) $m_3=56.5$ GeV,  
(b) $m_3=114$ GeV, and  (c) $m_3=228$ GeV. 
The each curve  has a left end point, in the left  region from which there is no real solution for $m_1/m_2$. 
The shaded square region is correspond to the observed 
mass ratios  in (4.11) for  up-quark sector obtained 
by Xing et. al.  
\end{quotation}
  \label{fig-up1}
\end{figure}

\begin{figure}[h!]
  \includegraphics[width=140mm,clip]{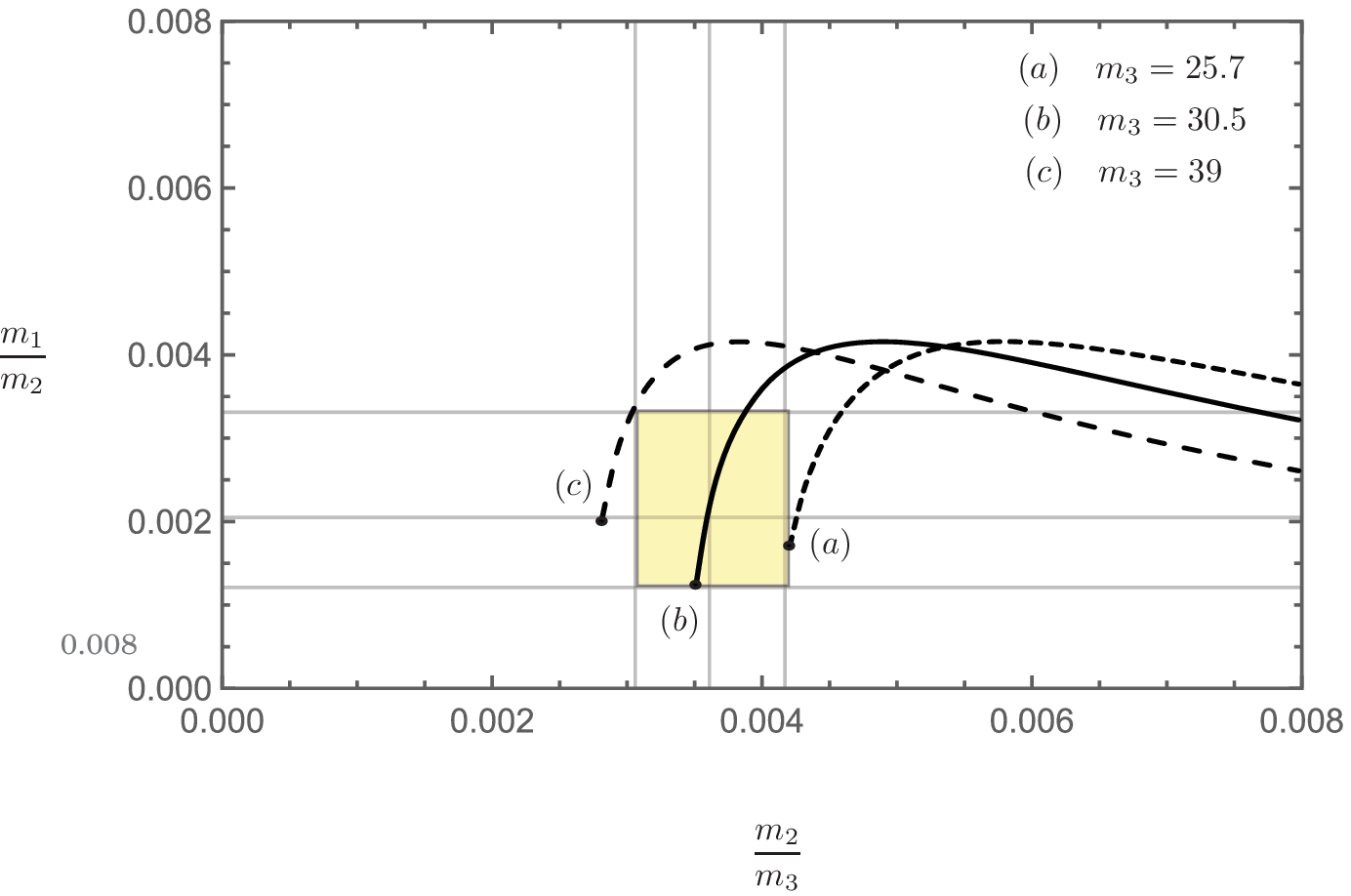}
\begin{quotation}
Figure 4: \,\,Another  $m_3$ value in the mass relation consistent 
with the observed values  $m_u/m_c$ and $m_c/m_t$ in the 
up-quark  sector. The curves of the mass relation $y=f(x)$ 
in Eq.(4.4) are drown in the $(m_2/m_3, m_1/m_2 )$ plane 
for the cases with (a) $m_3=25.7$ GeV,  (b) $m_3=30.5$ GeV, 
and  (c) $m_3=39$ GeV. 
The each curve  has a left end point, in the left  region from which there is no real solution for $m_1/m_2$. 
The shaded square region is correspond 
to the observed mass ratios  in (4.11) for  up-quark sector 
obtained by Xing et. al. 
\end{quotation}
  \label{fig-up2}
\end{figure}

\vspace{2mm}

\noindent{\bf Up-quark sector} 

In order to get a reasonable value of $m_3$ in the mass 
relation, we illustrate curves of the mass relation 
for several values of $m_3$ in Fig. 3 and Fig 4. 
We find  that there are two solutions of the $m_3$ 
which are consistent with the observed up-quark 
mass ratios (4.11) as seen in Figs.3 and 4: 
$$
m_3 = 30.5^{+8.5}_{-4.8}\ {\rm GeV} , \ \ \ \ \ 
m_3 =114^{+114}_{-\ 57}\ {\rm GeV} .
\eqno(4.16)
$$
Both solutions can give reasonable quark mass ratios
$(m_u/m_c, m_c/m_t) \simeq (2.4, 3.6)\times 10^{-3}$.
However, those two center values  $m_3=30.5$ GeV and 
$m_3 = 114$ GeV give 
$$
k_u= 5.63 \ \ \ {\rm and} \ \ \  k_u = 1.51,
\eqno(4.17)
$$ 
respectively. 
On the other hand, we have obtained $k_d =2.9$ in the 
down-quark sector as seen in Eq.(4.14). 
Therefore, whichever we take the value in (4.17), 
the value is poor agreement with $k_d=3$. 
It is natural to consider that the relations of quark 
mass matrices $M_u$ and $M_d$ to the charged lepton 
mass matrix $M_e$ take the same weight between the 
up- and down-quark sectors, i.e. $k_u = k_d$, 
except for the parameters $a_u$ and $a_d$.
If we want to consider $k_u = k_d = 3$,  
we have to choose $m_3= 57$ GeV from $m_t^{obs}=172$
GeV.  
Only when we chose the lowest value $m_3=57$ GeV 
 in the solution $m_3= 114^{+114}_{- \ 57}$ GeV, 
the value can give $k_u =3$, so that we can 
realize the relation $k_u = k_d =3$.
(If we require $k_u = k_d = 2.98$, the case leads 
to $m_3 = 59.4$ GeV, which is within the lower 
limit value $m_3 = 57$ GeV.) 

\vspace{2mm}

\noindent{\bf First generation quark masses} 

Next, we see the constraints on the first 
generation quark masses $m_u$ and $m_d$. 
From the curves (a) $m_3 = 0.91$ GeV, 
(b) $m_3 = 1.03$ GeV and (c) $m_3=1.04$ GeV 
in Fig.2,
we obtain 
$$
m_d = 2.9^{+3.7}_{-1.8} \ {\rm MeV}, 
\eqno(4.18)
$$
where we have used the input value \cite{q-mass} 
$(m_s)^{obs} = 55$ MeV.
Our result (4.18) has wide error compared 
with the observed value \cite{q-mass}
$(m_d)^{obs} = 2.90^{+1.24}_{-1.19}$ MeV. 

Similarly, from Figs.3 and 4, we obtain the following
two solutions of $m_u$, 
$$
m_u = 1.49^{+0.45}_{-0.75} \ {\rm MeV}, 
\ \ \ {\rm and} \ \ \ m_u =1.3^{+0.7}_{-0.5} \ 
{\rm MeV} ,
\eqno(4.19)
$$
respectively, 
where we have taken $(m_c)^{obs}= 0.619$ GeV.  
We also see that our results (4.19) cannot put
any severe constraint on the observed value 
$(m_u)^{obs} = 1.27^{+0.50}_{-0.42}$ MeV. 
\\


{\large\bf 5  \  Concluding remarks }  

In conclusion, we have investigate a pameter-independent
quark mass relation in the U(3)$\times$U(3)$'$ model.
Considering our results with the observed quark mass 
values \cite{q-mass}, we conclude that the choice 
$k_u=k_d=3$ in the previous work \cite{KN_PRD15} 
of the explicit parameter fitting of $a_u$ and $a_d$ 
was reasonable. 
However, we have found that there are two solutions in  
the up-quark sector as we have shown in Figs.3 and 4. 
This is not so serious problem when we take 
the error range of the observed quark mass 
values into consideration. 

We are convinced that our parameter-independent analysis of the mass relation is 
useful for model checking in future study of the 
U(3)$\times$U(3)$'$ model. 

We did not investigate a similar parameter-independent 
study for the CKM mixing. 
The similar study of the CKM matrix elements $V_{ij}$ 
cannot been obtained unless the results include quark 
masses. 
We are interested in the values $V_{us}$, $V_{cb}$, 
$V_{td}$ and so on, while those values will be disturbed 
by existence of large elements $V_{ud}$, $V_{cs}$, 
$V_{tb}$ and so on. 
It is our future task to obtain a relation without 
such large contribution terms.


\vspace{5mm}

One (Y.K.) of the authors  was supported by JSPS KAKENHI Grant
number JP16K05325.


\vspace{5mm}
%

\end{document}